\documentclass[twocolumn]{emulateapj}

\newcommand{\etal}{{\it et al.}}

\newcommand{\be}{\begin{equation}}
\newcommand{\ee}{\end{equation}}

\usepackage{pdfpages} 
 
\shorttitle{2MASS TF Template Erratum}
\shortauthors{Masters \etal}

\begin{document}
\title{Erratum: 2MTF I. The Tully-Fisher Relation in the 2MASS J, H and K-bands (2008 AJ 135, 1738)}
\author{Karen L. Masters\altaffilmark{1}, Christopher M. Springob\altaffilmark{2,3} \& John P. Huchra\altaffilmark{1,4}}
\altaffiltext{1}{Harvard-Smithsonian Center for Astrophysics, 60 Garden Street, Cambridge, MA 02138}
\altaffiltext{2}{Department of Physics and Astronomy, Washington State University, Pullman, WA 99164}
\altaffiltext{3}{Naval Research Laboratory, Remote Sensing Division Code 7213, 4555 Overlook Avenue, SW, Washington, D.C., 20375}
\altaffiltext{4}{JPH passed away in 2010, before the errors resulting in the need for this erratum were discovered}
\email{karen.masters@port.ac.uk}

\keywords{Erratum: 2MTF I. The Tully-Fisher Relation in the 2MASS J, H and K-bands (Masters, Springob and Huchra 2008 AJ 135, 1738). In this astroph version, the original paper follows in the same pdf}

\section{Need for Erratum}
 An sign error was noticed (by Hong Tao) in the application of the internal extinction and k-corrections to magnitudes used in Masters et al. (2008). Both mistakes meant that the corrections were positive (i.e. dimming the magnitudes) rather than negative. The majority of reported magnitudes from that work are therefore dimmer than they should be. However, for a small number of very low luminosity galaxies the corrected magnitudes were brighter than they should be (the smallest correction was erroneously allowed to drop below zero).
  
\begin{figure*} 
\includegraphics[width=16cm,bb=92 352 551 772]{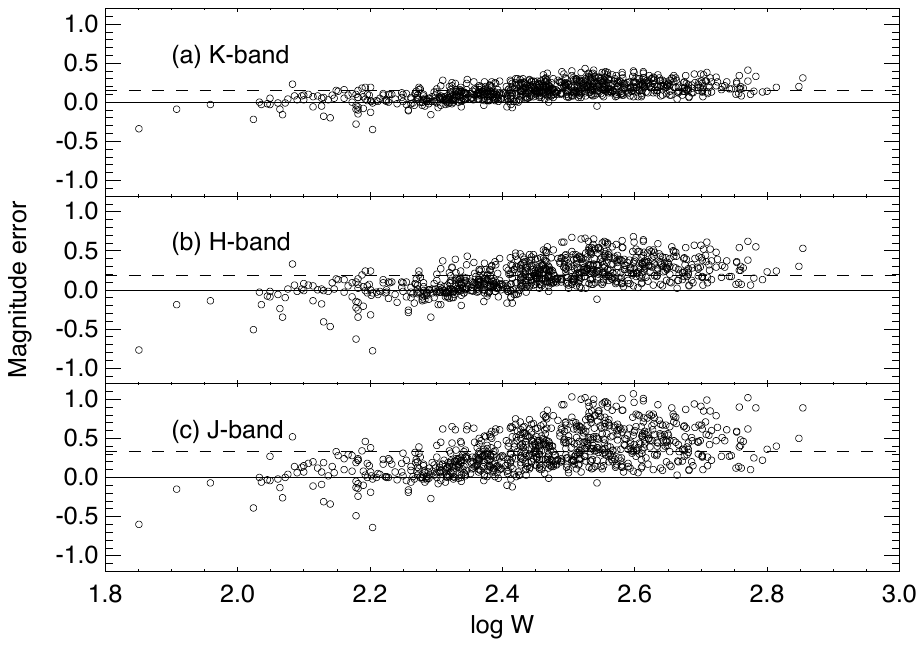}
\caption{Magnitude errors plotted as a function of rotation width. The offset can be larger for galaxies with large $\log W$ which have larger internal extinction corrections if they are very inclined. For low luminosity (width) galaxies the offset can be negative. The average error across the whole sample is shown as the dashed line. 
\label{errors}}
\end{figure*}

 The average magnitude error over all $N=888$ galaxies of the sample are 0.34 magnitudes in J-band, 0.18 magnitude in H-band and 0.15 magnitudes in K-band.  We show the size of the error for all galaxies in the sample in Figure \ref{errors}. Unfortunately because internal extinction depends on galaxy luminosity, the size of the error varies along the TF relation, so both the slope and the offset are subtly affected.  This error propagated through the entire analysis of Masters et al. (2008), changing all quantitative results. We have rerun the entire analysis and publish corrected numbers and data tables for that work here.  

\subsection{Updated Morphological Correction}
 There is no qualitative change in the TF morphological corrections (Section 3 point 5 in Masters et al. 2008). The updated values are small fractions of a magnitude different from the original values published. They are:  
 K band:
\begin{displaymath}
\begin{array}{lllr}
\Delta M_{\rm Sa} & = & 0.27 - 3.73 (\log W - 2.5): & T \leq 2   \\
\Delta M _{\rm Sb} & = & 0.02 - 1.74 (\log W - 2.5): & 3 \leq T \leq 4 ,
\end{array}
\end{displaymath}

H band:
\begin{displaymath}
\begin{array}{lllr}
\Delta M_{\rm Sa} & = & 0.24 - 3.60 (\log W - 2.5): & T \leq 2  \\
\Delta M _{\rm Sb} & = & 0.01 - 1.72 (\log W - 2.5): & 3 \leq T \leq 4
\end{array}
\end{displaymath} 

J band: 
\begin{displaymath}
\begin{array}{lllr}
\Delta M_{\rm Sa} & = & 0.23 - 3.69 (\log W - 2.5): & T \leq 2  \\
\Delta M _{\rm Sb} & = & 0.00 - 1.91 (\log W - 2.5): & 3 \leq T \leq 4.
\end{array}
\end{displaymath}

\subsection{Incompleteness Correction}
 
 Incompleteness bias corrections depend on the amount of incompleteness in a cluster sample, and the input TF and scatter. 
 
 For the updated incompleteness functions, we use the shapes of the functions derived in Masters et al. (2008), but correct them by the average offset in magnitude (due to the error) for each individual cluster subsample.  
  
 The input TF relations have been updated to the below (these are corrected with the galaxy morphology corrections as above) 
\begin{eqnarray}
M_K - 5\log h & = & -22.396 - 9.484 (\log W - 2.5), \nonumber \\
M_H - 5\log h & = & -22.182 - 9.260 (\log W - 2.5), \nonumber \\
M_J - 5\log h & = & -21.597 - 9.230 (\log W - 2.5). \nonumber 
\end{eqnarray} 

 The scatter we use is the scatter from the above relations which are: 
\begin{eqnarray}
\epsilon_{\rm int, K} &= &0.42 - 1.63 (\log W - 2.5), \nonumber \\
\epsilon_{\rm int, H} &= &0.46 - 1.53 (\log W - 2.5), \nonumber \\
\epsilon_{\rm int, J} &= &0.47 - 1.39 (\log W - 2.5), \nonumber
\end{eqnarray}
 
 The magnitude of the bias correction for a given galaxy depends the absolute magnitude of the galaxy. As these in general have brightened, the incompleteness is slightly larger than previously, and most incompleteness bias corrections have increased very slightly. On average the bias correction is larger by 0.02, 0.09 and 0.03 mag in K, H and J-bands respectively, with a slight trend for larger absolute changes in the correction for smaller width galaxies. 

\subsection{Final template}
The small changes carry through to the final template as below:  
\begin{eqnarray}
M_K - 5\log h & = & -22.188\pm0.015 - 10.74\pm0.10 (\log W - 2.5), \nonumber \\
M_H - 5\log h & = & -21.951\pm0.017 - 10.65\pm0.11 (\log W - 2.5), \nonumber \\
M_J - 5\log h & = & -21.370\pm0.018 - 10.61\pm0.12 (\log W - 2.5). \nonumber 
\end{eqnarray} 
This results in a template which in K-band is both brighter by 0.15 mag and steeper by 0.7 mag/dex. Both of these are statistically significant changes -- at about the 5-7$\sigma$ level. In H-band the change results in a 0.12 mag brightening of the zeropoint and a steepening of 1.6 mag/dex. In J-band we see a 0.37 mag brightening of the zero point and a steepening of 1.5 mag/dex. 

An updated version of table 2 from Masters et al. (2008) providing a summary of all TF fits to the corrected data is provided below in Table \ref{globalfits}. 

\newpage
\subsection{Intrinsic Scatter}
The new template results in only minor changes to the measured scatter. The updated observed scatter from the global TF in all three bands is shown below. These have zeropoints 0.01-0.04 mag larger than previously seen and also show a slightly steeper dependence on $\log W$. 
 \begin{eqnarray}
\epsilon_{\rm obs, K} &= &0.56 - 0.92 (\log W - 2.5)  \nonumber \\
 \epsilon_{\rm obs, H} &= & 0.58 - 0.96  (\log W - 2.5)\nonumber \\
 \epsilon_{\rm obs, J} &= &0.60 - 0.80 (\log W - 2.5) \nonumber
\end{eqnarray}
Once corrected for measurement errors this results in an intrinsic scatter of: 
 \begin{eqnarray}
 \epsilon_{\rm int, K} &= 0.34 -  1.31 (\log W - 2.5) \nonumber \\
\epsilon_{\rm int, H} &=  0.38 - 1.28  (\log W - 2.5) \nonumber \\
\epsilon_{\rm int, J} &= 0.40 - 1.05 (\log W - 2.5) \nonumber 
\end{eqnarray}

\begin{deluxetable}{lccccccccccc}
\tablecolumns{12} 
\tablewidth{0pc} 
\tabletypesize{\tiny}
\tablecaption{TF Fit Parameters \label{globalfits}}
\tablehead{ 
\colhead{Sample} & \colhead{$N$} & \multicolumn{2}{c}{Direct fit} & \multicolumn{2}{c}{Inverse fit}  & \multicolumn{5}{c}{Bivariate fit}   \\
\colhead{} & \colhead{} & \colhead{$a_{dir}$ }    & \colhead{$b_{dir}$} & 
\colhead{$a_{inv}$ } & \colhead{ $b_{inv}$}   & \colhead{$a_{bi}$ }    & \colhead{$b_{bi}$} & 
\colhead{$\epsilon_a$ }    & \colhead{$\epsilon_b$}   & \colhead{$\sigma_{\rm SD}$ }    & \colhead{$\sigma_{\rm abs}$ }
} 
\startdata
\cutinhead{\bf K-band}
\sidehead{No corrections}
      Full & 888& -22.439&  -6.650& -22.435& -10.620& -22.385&  -7.500&   0.020&   0.116&   0.589&   0.413\\
     Sa &   101& -22.759&  -5.104& -22.490&  -7.773& -22.673&  -5.527&   0.064&   0.320&   0.504&   0.372\\
    Sb &   345& -22.468&  -6.631& -22.402&  -9.933& -22.423&  -7.518&   0.030&   0.190&   0.583&   0.391\\
Sc   &   442& -22.542&  -7.612& -22.619& -12.348& -22.403&  -9.255&   0.022&   0.176&   0.516&   0.371\\
\sidehead{Morphological correction only}
Full &   888& -22.485&  -8.215& -22.424& -10.604& -22.396&  -9.484&   0.018&   0.106&   0.520&   0.354\\
\sidehead{Bias correction only}
Sa &   101& -22.537&  -6.176& -22.476&  -7.810& -22.441&  -6.652&   0.060&   0.303&   0.492&   0.364\\
Sb &  345& -22.229&  -8.160& -22.395&  -9.936& -22.200&  -9.142&   0.027&   0.185&   0.588&   0.418\\
Sc &   442& -22.300&  -8.882& -22.608& -12.309& -22.204& -10.793&   0.020&   0.172&   0.541&   0.402\\
Full & 888& -22.201&  -8.110& -22.426& -10.608& -22.161&  -9.044&   0.018&   0.114&   0.614&   0.448\\
\sidehead{All corrections}
Full &   888& -22.278&  -9.170& -22.418& -10.597& -22.188& -10.736&   0.015&   0.100&   0.526&   0.377\\
Low mass & 303& -21.989&  -9.169\\
High mass &  374& -22.514&  -8.099\\
\cutinhead {\bf H-band}
\sidehead{No corrections}
Full &   888& -22.196&  -6.790& -22.207& -10.655& -22.159&  -7.591&   0.020&   0.130&   0.617&   0.436\\
Sa &   101& -22.514&  -5.158& -22.237&  -7.922& -22.430&  -5.582&   0.066&   0.345&   0.520&   0.396\\
Sb &   345& -22.228&  -6.626& -22.173& -10.014& -22.195&  -7.466&   0.032&   0.213&   0.624&   0.425\\
Sc & 442& -22.265&  -7.469& -22.396& -12.373& -22.186&  -9.181&   0.023&   0.190&   0.525&   0.375\\
\sidehead{Morphological correction only}
   Full &  888& -22.227&  -8.236& -22.198& -10.651& -22.182&  -9.260&   0.019&   0.120&   0.564&   0.394\\
\sidehead{Bias correction only}
Full &   888& -21.349&  -8.381& -21.608& -10.593& -21.338&  -9.173&   0.020&   0.137&   0.660&   0.485\\
Sa &   101& -22.270&  -6.355& -22.218&  -7.978& -22.176&  -6.840&   0.061&   0.320&   0.514&   0.390\\
Sb &   345& -21.976&  -8.159& -22.167& -10.003& -21.959&  -9.064&   0.028&   0.198&   0.612&   0.434\\
Sc &   442& -22.012&  -8.808& -22.386& -12.354& -21.971& -10.816&   0.022&   0.187&   0.552&   0.408\\
\sidehead{All corrections}
Full &   888& -21.992&  -9.316& -22.191& -10.648& -21.951& -10.648&   0.017&   0.113&   0.561&   0.405\\
Low mass &   303& -21.690&  -8.838\\ 
High mass &  374& -22.213&  -8.219\\
\cutinhead{\bf J-band}
\sidehead{No corrections}
Full &   888& -21.591&  -6.794& -21.617& -10.598& -21.565&  -7.563&   0.021&   0.141&   0.644&   0.456\\
Sa &    101& -21.909&  -5.081& -21.642&  -7.873& -21.827&  -5.501&   0.069&   0.362&   0.525&   0.399\\
Sb &    345& -21.619&  -6.531& -21.573&  -9.999& -21.601&  -7.289&   0.035&   0.242&   0.665&   0.455\\
Sc &   442& -21.617&  -7.301& -21.814& -12.311& -21.600&  -9.195&   0.024&   0.200&   0.565&   0.396\\
\sidehead{Morphological correction only}
Full &   888& -21.625&  -8.299& -21.608& -10.591& -21.597&  -9.230&   0.020&   0.129&   0.608&   0.424\\
\sidehead{Bias correction only}
Full &   888& -21.349&  -8.375& -21.608& -10.593& -21.337&  -9.167&   0.020&   0.137&   0.660&   0.485\\
Sa&  101& -21.666&  -6.291& -21.625&  -7.930& -21.577&  -6.770&   0.062&   0.335&   0.528&   0.406\\
Sb &   345& -21.374&  -8.027& -21.566&  -9.996& -21.371&  -8.898&   0.032&   0.225&   0.647&   0.462\\
Sc &  442& -21.372&  -8.663& -21.805& -12.286& -21.389& -10.716&   0.023&   0.194&   0.594&   0.431\\
\sidehead{All corrections}
Full &   888& -21.384&  -9.597& -21.601& -10.588& -21.370& -10.612&   0.018&   0.124&   0.611&   0.441\\
Low mass & 303& -21.069&  -8.530&\\
High mass &  374& -21.561&  -8.554\\
\enddata
\tablecomments{Column (1) describes the sample used to fit to, column (2) shows the total number of galaxies in that sample. Columns (3-8) show parameters for $M - 5 \log h = a + b (\log W - 2.5)$ using direct, inverse and bivariate fits. Also shown for the bivariate fits are the error on $a$ and $b$ (columns 9 \& 10), and the scatter calculated as the standard deviation (column 11) and absolute deviation (column 12). Inverse and bivariate fits are not reported for the sample divided by galaxy size since the artificial $\log W$ cut-off which is applied creates a large bias in those fits.}
\end{deluxetable}

\begin{acknowledgments}
We apologise for any inconvenience caused by this error. We thank Tao Hong for his careful work which caused him to identify it. 
\end{acknowledgments}

\newpage
\includepdf[pages={1,{},2-13},scale=0.95]{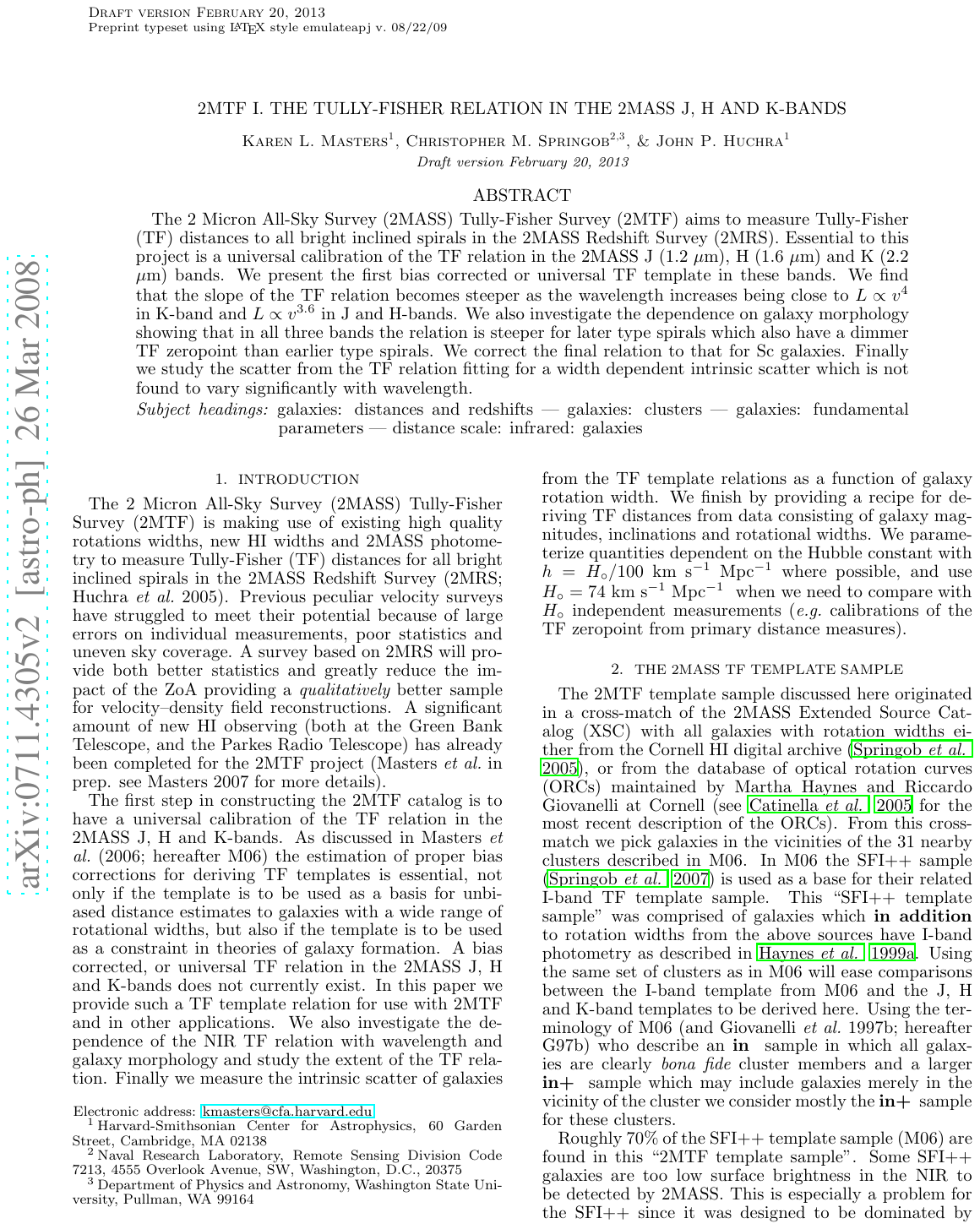}

\end{document}